\documentclass{article} % For LaTeX2e
\usepackage{colm2024_conference}

\usepackage{microtype}
\usepackage{url}
\usepackage{booktabs}
\usepackage[utf8]{inputenc} % allow utf-8 input
\usepackage[T1]{fontenc}    % use 8-bit T1 fonts
\usepackage{hyperref}       % hyperlinks
\usepackage{url}            % simple URL typesetting
\usepackage{booktabs}       % professional-quality tables
\usepackage{multirow}

\usepackage{amsfonts}       % blackboard math symbols
\usepackage{nicefrac}       % compact symbols for 1/2, etc.
\usepackage{microtype}      % microtypography
\usepackage{xcolor}         % colors
\usepackage{amssymb, amsmath}
\usepackage{graphicx}
\usepackage{amsmath}
\usepackage{float}
\usepackage{enumitem}
\usepackage{float}
\usepackage{algorithm}
\usepackage{algpseudocode}
\setlist[itemize]{leftmargin=*}
\newtheorem{definition}{Definition}

\usepackage{hyperref}

\usepackage{hyperref}
\hypersetup{
    hidelinks,
    breaklinks,
    colorlinks,
    linkcolor={blue!70!black},
    citecolor={red!70!black},
    urlcolor={blue!70!black}
}

\title{ProLLM: Protein Chain-of-Thoughts Enhanced LLM
for Protein-Protein Interaction Prediction}

\author{%
  Mingyu Jin\thanks{Equal Contribution.} \\
  Rutgers University \\
  \And
  Haochen Xue$^{*}$\\
  University of Liverpool \\
   \And
  Zhenting Wang \\
  Rutgers University \\
  \And
  Boming Kang\\
  Peking University \\
  \And
  Ruosong Ye \\
  Rutgers University \\
  \And
  Kaixiong Zhou \\
  Massachusetts Institute of Technology \\
  \And
  Mengnan Du \\
  New Jersey Institute of Technology \\
  \And
  Yongfeng Zhang\thanks{Author Emails: mingyu.jin@rutgers.edu, hicaca945@gmail.com, zhenting.wang@rutgers.edu, kangbm@hsc.pku.edu.cn,  ruosong.ye@rutgers.edu,  kz34@mit.edu, mengnan.du@njit.edu, yongfeng.zhang@rutgers.edu} \\
  Rutgers University \\
}

\colmfinalcopy % Uncomment for camera-ready version, but NOT for submission.
\begin{document}

\maketitle

\begin{abstract}
The prediction of protein-protein interactions (PPIs) is crucial for understanding biological functions and diseases. Previous machine learning approaches to PPI prediction mainly focus on direct physical interactions, ignoring the broader context of nonphysical connections through intermediate proteins, thus limiting their effectiveness. The emergence of Large Language Models (LLMs) provides a new opportunity for addressing this complex biological challenge. By transforming structured data into natural language prompts, we can map the relationships between proteins into texts. This approach allows LLMs to identify indirect connections between proteins, tracing the path from upstream to downstream. Therefore, we propose a novel framework \textbf{ProLLM} that employs an LLM tailored for PPI for the first time. Specifically, we propose \textbf{Protein Chain of Thought (ProCoT)}, which replicates the biological mechanism of signaling pathways as natural language prompts. ProCoT considers a signaling pathway as a protein reasoning process, which starts from upstream proteins and passes through several intermediate proteins to transmit biological signals to downstream proteins. Thus, we can use ProCoT to predict the interaction between upstream proteins and downstream proteins. The training of ProLLM employs the ProCoT format, which enhances the model's understanding of complex biological problems. In addition to ProCoT, this paper also contributes to the exploration of embedding replacement of protein sites in natural language prompts, and instruction fine-tuning in protein knowledge datasets. We demonstrate the efficacy of ProLLM through rigorous validation against benchmark datasets, showing significant improvement over existing methods in terms of prediction accuracy and generalizability. Our results highlight the potential of LLMs to transform the field of PPI, serving as a robust potential tool for various categories of biological and medical research. The code is available at: \url{https://github.com/MingyuJ666/ProLLM}.
\end{abstract}

\section{Introduction}
Protein-protein interactions (PPIs) play an essential role in various biological processes of all living organisms, which are crucial for biomedical, genetic, and pharmaceutical research. Thus, numerous experimental methods have been proposed for PPI detection, such as yeast two-hybrid~\citep{ito2001comprehensive} and quantitative proteomics methods~\citep{rotilio2012proteomics}. However, wet-lab methods for PPI prediction are often time-consuming and labor-intensive, highlighting the need for more precise and efficient computational tools.

In recent years, computational biology has developed rapidly. Methods such as the Convolutional Neural Network (CNN) and Graph Neural Network (GNN) have become powerful tools for studying protein interaction. CNN-based approaches like TAG-PPI~\citep{song2022learning} typically use pre-trained embedding models to convert protein sequences into numerical vector representations, and then employ one dimensional convolutional neural networks to extract features from the vectors for subsequent PPI tasks. 

Although CNN methods have shown some effectiveness in PPI prediction, they still have limitations due to their fixed receptive fields and the lack of well-defined spatial relationships in protein sequences, which limit the accuracy and interpretability of the predictions. GNN-based methods such as GNN-PPI~\citep{lv2021learning} treat proteins as nodes and their relationships as edges, constructing a network composed of proteins, which better captures protein relationships and interactions, and outperforms CNNs in predicting protein interactions. However, while GNNs can effectively extract network structural information, they neglected the non-physical connections between two proteins without direct physical interactions, resulting in the worse performance in learning protein chains than transformer-based models ~\citep{zhou2020graph}. Furthermore, GNNs cannot fully capture the relationships and dynamic changes in the signal passing process of living organisms, restricting their performance for PPI prediction ~\citep{zhou2023protein}.

% add why 

\begin{figure}[]
\centering
\vspace{-0.8cm}
\includegraphics[width=0.9\linewidth]{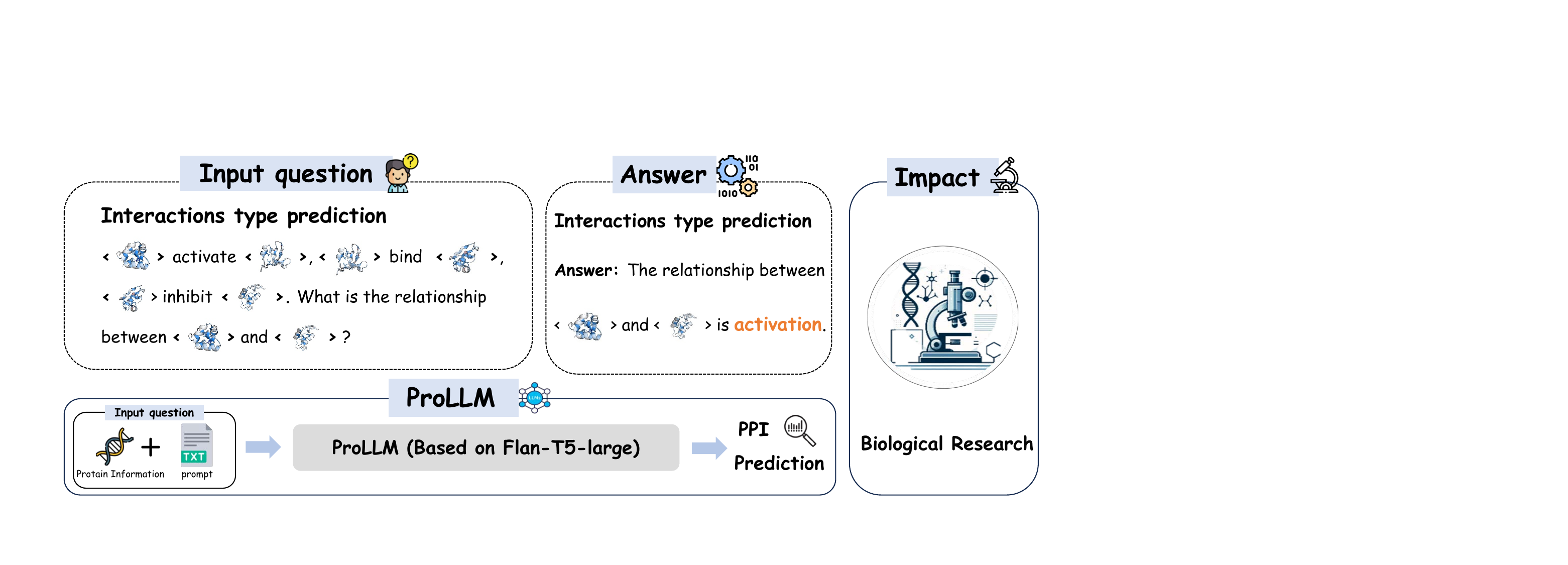}
\caption{Illustration of the ProLLM Framework. We fine-tuning ProLLM under Human, SHS27K, SHS148K, and STRING datasets, enabling it to solve various PPI related tasks with the structure information purely described by natural language.}
\label{fig:image1}
\end{figure}

% q1 q2 合并就不是问题了

Following the GNN and CNN methods, Large Language Models have also been applied to this PPI area, such as ProBert~\citep{elnaggar2021prottrans} and ProteinLM~\citep{xiao2021modeling}. As long as these models can obtain a protein representation, we can use the direct cosine similarity of the representation or train an MLP to perform PPI prediction. However, these methods still cannot capture the chain relationships between proteins, such as the signaling pathways. Besides, previous literature only used LLMs as a feature extractor. Recently, using LLM as a link predictor has shown that it can better capture relational information between nodes in knowledge graph tasks and its performance surpasses traditional GNN baselines~\citep{ye2023language,zhuo2024protllm,shu2024knowledge}. Therefore, it is promising to introduce LLM for protein-protein interaction (PPI) tasks, since the most important biological signal for PPI tasks is the chain relationships of proteins, i.e., the signaling pathways.

\begin{figure}
\centering
\vspace{-0.8cm}
\includegraphics[width=0.9\linewidth]{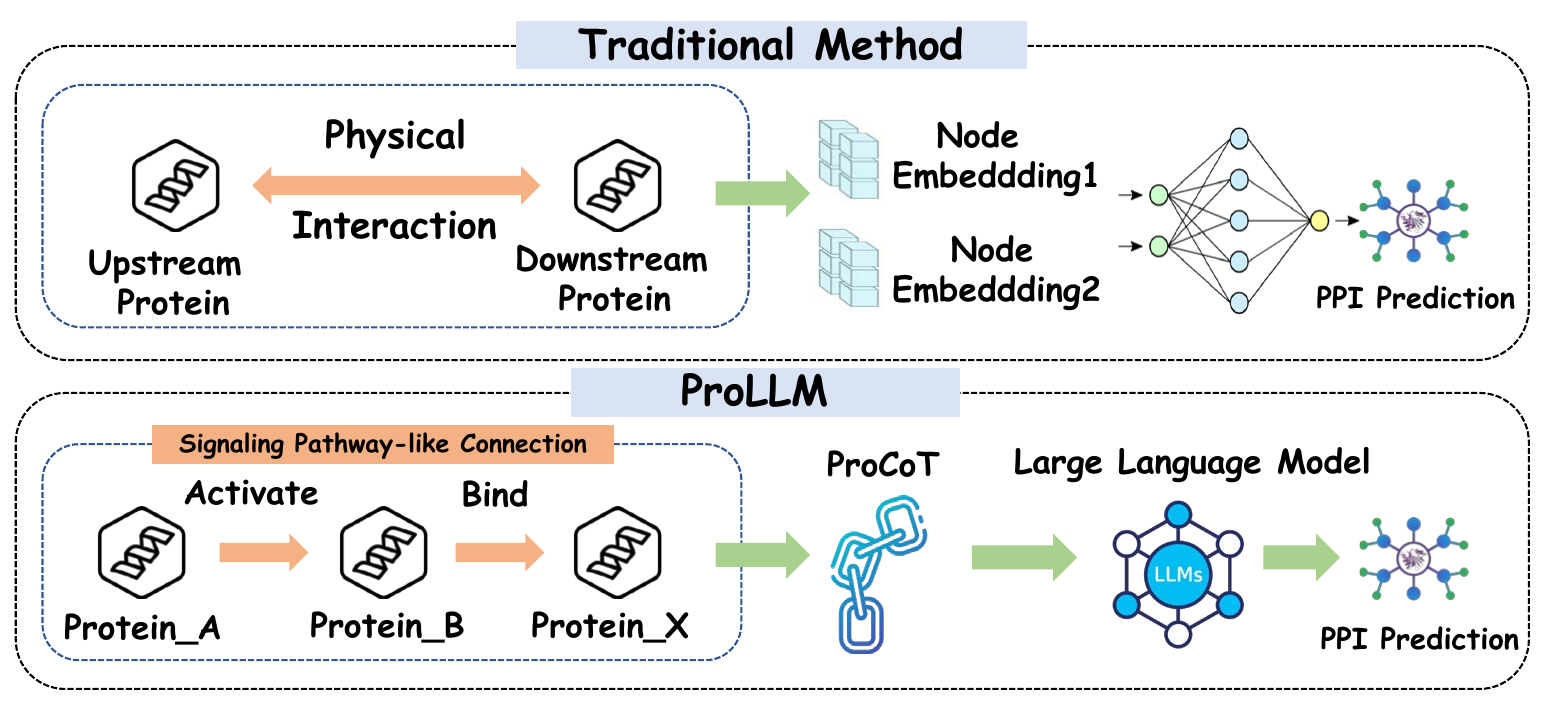}
\caption{The difference between the existing method and our method in PPI prediction. Existing method focus on the property of upstream protein and downstream protein, our method focus on signaling pathway-like connection.}
\label{fig:compare}
\end{figure}

To bridge the gap, we propose \textbf{ProLLM}, with its key ideas illustrated in \autoref{fig:image1}, and the difference between the existing method and our ProLLM shown in \autoref{fig:compare}. Existing methods only focus on the single protein-protein interaction, overlooking the application of protein links to predict PPI in signaling pathways. Instead, we employ a large language model to learn the law of signal pathways and adapt the LLM to directly predict the type of interaction between proteins. 

The signaling pathway addresses the traditional method's ignorance of global, non-physical connections between proteins. Signaling pathways typically start with an upstream protein that sends a biological signal through several intermediates to a downstream protein, hence requiring consideration of the cumulative effect of multiple protein interactions. This series of interactions form sequential chains. Therefore, we propose \textbf{Protein Chain of Thought (ProCoT)} to overcome the limitation in understanding signaling pathways and protein functions. ProCoT is a data format that simulates the signal transduction process using a thought-chain approach, thereby enabling the prediction of protein interactions in signaling pathway problems. CoT can express the thinking process step by step to form a reasoning chain~\citep{jin2024impact}, while ProCoT extends this principle further into the protein-related domain to simulate protein signaling pathways, giving LLMs a deeper insight into protein.

Additionally, our approach addresses the issue of poor protein comprehension in LLMs by replacing the standard language model embedding with embedding infused with protein information. When we process the protein name in the prompt, we replace its original embedding by ProtTrans~\citep{9477085}, because its embedding contains protein's structual information. We also perform the instruction fine-tuning on the protein knowledge dataset to infuse protein domain knowledge into LLMs. Following these steps, the LLM acquires a robust ability to reason about the direct relationships between proteins, as demonstrated in \autoref{fig:image1}. It can provide answers to questions about protein relationships, which play a significant role in biological research.

Our contributions are summarized as follows:
\begin{itemize}
    \item 
    To the best of our knowledge, we are the first to explore the PPI prediction problem as a natural language processing problem. The proposed Protein Chain of Thought (ProCoT) is a novel method for understanding complex multi-step protein interactions, i.e., those present in signaling pathways.

    \item We propose embedding replacement and instruction fine-tuning on our model to enhance its ability to understand and reason about proteins. This also provides our model with rich background knowledge of protein sequences and protein interactions before training.

    \item Experiments on four widely used PPI datasets (i.e., Human, SHS27K, SHS148K, and STRING) demonstrate that ProLLM outperforms graph-based methods such as GNN-PPI and SemiGNN-PPI. It also has better performance than LLM-based methods like InstructGLM. The micro-F1 scores on these 4 datasets are 91.05, 78.09, 87.66, 89.21, respectively.

\end{itemize}

\section{Related Work}
\paragraph{Protein-protein Iteractions}

Protein-protein interactions (PPIs) are indispensable for all living organisms, and so many efforts have been made for PPI detection up to now. Yeast two-hybrid (Y2H) assays~\citep{bruckner2009yeast}, synthetic lethality~\citep{o2017synthetic},  quantitative proteomics~\citep{wilm2009quantitative}, and mass spectrometry~\citep{mann2001analysis}, are widely used for identifying PPIs. To be specific, Y2H assays explore binary PPIs in living cells, offering insight into protein functions and regulations, although labor-intensive and limited in genomic coverage. Synthetic lethality identifies essential gene pair interactions by revealing compensatory relationships when both are inactivated. Meanwhile, quantitative proteomics, especially through mass spectrometry, illuminates the dynamic nature of the interactome under various conditions. Although informative, these methods require significant labor, time, and resources.

\paragraph{Traditional Machine Learning Models for PPI Prediction}

In the realm of machine learning, sequence-based approaches include Shen's SVM method~\citep{shen2007predicting}, which uses a 3-mer count vector from protein sequences as features and groups 20 amino acids into seven classes to handle synonymous mutations and reduce feature space dimensionality. SVM-based methods~\citep{guo2008using} and the ensemble model PCA-EELM (Principal Component Analysis-Ensemble Extreme Learning Machine)~\citep{you2013prediction} utilize various types of protein sequence information for PPI prediction.
In the domain of deep learning, DeepPPI~\citep{du2017deepppi} extracts a multitude of features from protein sequences and employs a dual deep neural network structure for feature fusion and prediction. Sun et al.~\citep{sun2017sequence} introduced a PPI predictor based on a stacked autoencoder, emphasizing the importance of sample balance. DPPI~\citep{hashemifar2018predicting} and TAGPPI~\citep{song2022learning} further extend the application of convolutional neural networks and integrate text convolution networks with graph representation learning to enhance the accuracy of PPI predictions.
GNNs have significantly advanced PPI predictions, improving our understanding of biological mechanisms. GNN-PPI~\citep{lv2021learning} enhances inter-novel-protein prediction accuracy by utilizing protein relationships and a new evaluation approach. PT-GNN~\citep{long2022pre} integrates diverse data for link prediction, learning node features from sequence and structure. DeepRank-GNN~\citep{reau2023deeprank} offers a modular, Python-packaged framework for GNN-based interaction pattern predictions. HIGH-PPI~\citep{gao2023hierarchical} introduces a hierarchical graph learning model for effective PPI prediction and molecular detail extrapolation.
Geometric GNNs excel in modeling spatial intricacies, enhancing biomolecule prediction accuracy. Geo-PPI~\citep{liu2021deep} utilizes self-supervised learning for geometric protein structure representations, excelling in detailing protein interactions. mmCSM-PPI~\citep{rodrigues2021mmcsm} captures multifaceted features for mutation impact predictions on protein interactions. MAPE-PPI~\citep{wu2024mape} defines the microenvironment of amino acid residues in proteins and encodes it into discrete codes, which can capture the connections between different microenvironments, enhancing the prediction accuracy of PPI.

\vspace{-0.3cm}
\paragraph{Large Language Model for PPI prediction}
Recent advances in large language models, such as BERT~\citep{BERT}, GPT~\citep{gpt4}, LLaMA~\citep{touvron2023llama}, and T5~\citep{T5}, have significantly advanced the field of Natural Language Processing (NLP) to new heights. These models, having been trained on extensive textual corpora, exhibit exceptional capabilities across a diverse range of NLP applications~\citep{shengyuan2024differentiable,jin2024health,fan2024nphardeval4v,hua2024disentangling}. Inspired by LLMs, Protein Large Language Models(PLMs) pre-trained on large-scale protein sequences have emerged, such as ESM~\citep{hsu2022learning}, ProtTrans~\citep{9477085} and ProteinBert~\citep{9477085}. PLMs provide a better representation of protein sequences by converting the protein sequences into the form of high-dimensional numerical vectors, known as embedding. With the protein sequences captured by the PLMs, the performances on diverse downstream tasks, such as structure prediction~\citep{lin2023evolutionary}, subcellular localization prediction~\citep{thumuluri2022deeploc}, single peptide prediction~\citep{teufel2022signalp} and N-linked glycosylation sites prediction~\citep{hou2023emngly}, have been transformed. It can be expected that PLMs will assist in PPI prediction tasks. ProtLLM~\citep{zhuo2024protllm} utilizes a dynamic protein mounting mechanism, a protein-as-word language modeling approach, and the InterPT dataset for pre-training, enabling it to handle complex inputs and achieve superior performance on various proteins-related tasks. However, ProtLLM is used for general protein tasks, and it is not used for PPI tasks exactly. The methodology of training these LLMs to convert text inputs to desired text outputs positions them as particularly advantageous for tasks such as generative link prediction~\citep{ye2023language, shu2024knowledge}. In such tasks, the model is tasked with inferring and generating the relationship between two entities based on provided textual cues. Moreover, the extensive pre-training phase enables LLMs to exhibit a remarkable capacity for generalization. This capacity allows them to effectively tackle and respond to tasks or prompts that were not explicitly covered during their initial training~\citep{wei2022emergent}. In addition to the outlined capabilities, the inherent flexibility and generalization potential of LLMs suggests that their applicability extends well beyond the conventional boundaries of NLP tasks~\citep{yang2023harnessing}. Specifically, their proficiency in generalizing from expansive pre-training sessions paves the way for their application in fields like bioinformatics and complex network analysis. 

\section{Proposed ProLLM Framework}
\begin{figure}
\centering
\vspace{-0.8cm}
\includegraphics[width=1.0\linewidth]{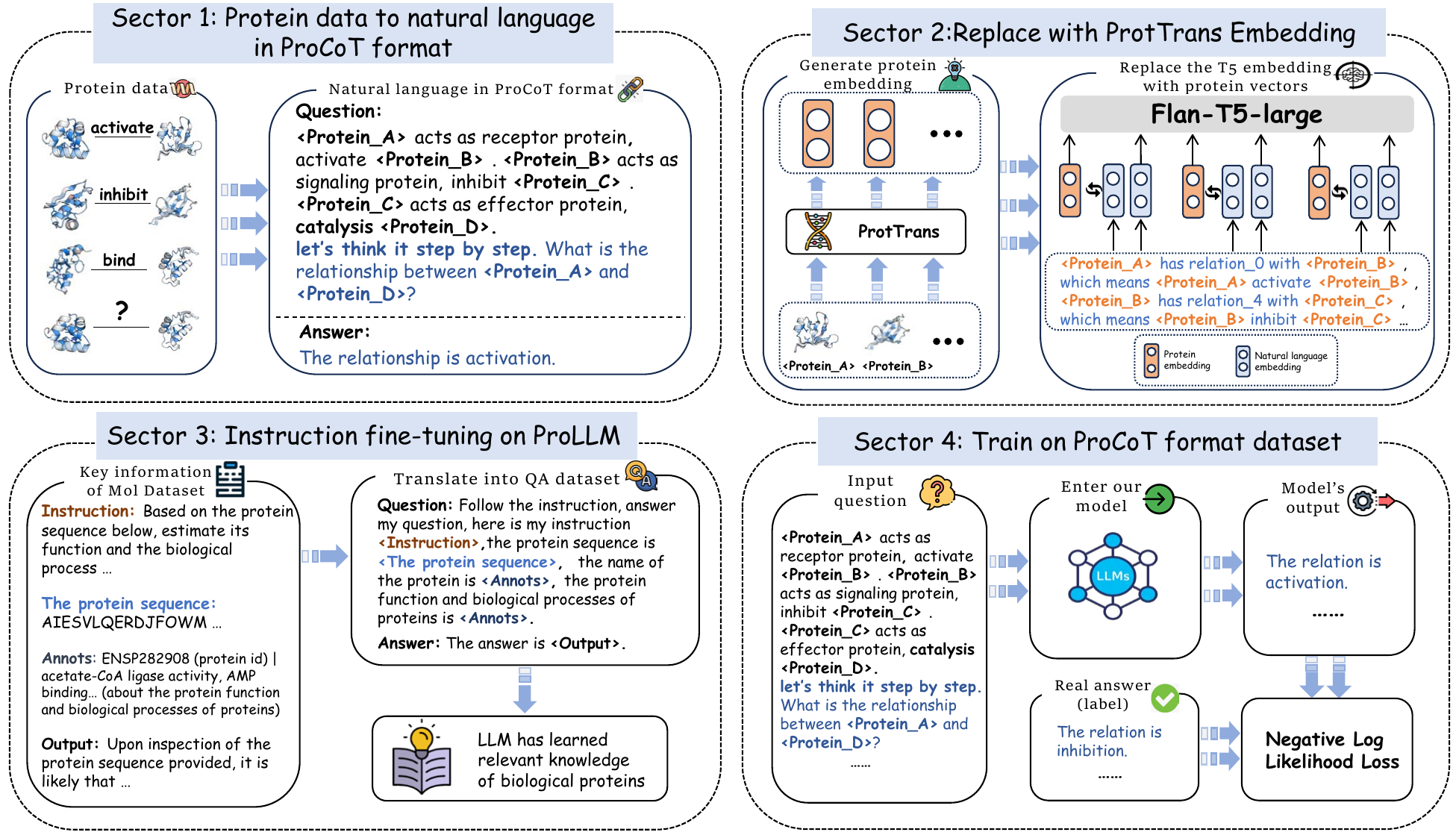}
\caption{The process of ProLLM. Sector 1: Transfer the original protein data into ProCoT format of natural language that indicates the signaling pathways between proteins; Sector 2: Replace protein information embeddings with natural language embeddings to enhance the model's understanding of proteins; Sector 3: Inject knowledge about protein function; Sector 4: Fine-tuning on the ProCoT format dataset in Sector 1.}
\label{fig:process}
\end{figure}

In this section, we will introduce the implementation details of \textbf{ProLLM}, a framework designed to transform protein interaction data into ProCoT format natural language descriptions to simulate protein signaling pathways. By translating the structure relationships of proteins into natural language, we effectively transform protein interaction problems into natural language processing (NLP) tasks. To enhance the understanding of proteins by ProLLM, we directly integrate the protein vectors generated from ProtTrans~\citep{9477085} into our model to replace the original word embedding in the protein name's place. This approach allows our model to understand and utilize the biological attributes of proteins when predicting protein interactions. The ProLLM process is shown in \autoref{fig:process}.

\subsection{ProCoT}
\subsubsection{Protein Data to Natural Language in CoT Format}
\label{sec:pcot}

Transformation of the original protein data into a ProCoT natural language is a critical step, this kind of natural language type that represents the relationships of proteins is shown as \autoref{def:procot}. 

\begin{definition}{(Protein Interaction)}\label{def:procot}
Given the set \( \mathcal{P} \) representing all proteins under consideration in the study, we define the interaction between any two proteins \( p_n, p_m \in \mathcal{P} \) with the specific type interaction \( R_i \) interaction set \( \mathcal{R} \). This relationship is encapsulated as a tuple \( (p_n, p_m, R_i) \). 
\end{definition}

\begin{definition}{(Protein Signaling Pathway)}\label{def:signal}
In cellular signal transduction, receptor protein \(\mathcal{R}\text{c}\) is the beginning of the signal pathway, and it can interact with downstream signaling proteins \(\mathcal{S}\text{p}\). \(\mathcal{S}\text{p}\) is responsible for continuing to transmit signals to effector protein \(\mathcal{E}\text{p}\) inside the cell. After the \(\mathcal{E}\text{p}\) response to the signal, it will trigger specific biological effects and activate secondary messengers \(\mathcal{M}\). \(\mathcal{M}\) may be the end of this signal level, or it may open a second signal hierarchy. If in the second signal level, the \(\mathcal{M}\) will replace the  \(\mathcal{R}\text{c}\) in the initial signal hierarchy as the starting point. The signal will continue to propagate from \(\mathcal{R}\text{c}\) through new signaling proteins, and effector proteins, and then generate new secondary messengers. After this stage, the propagation of the signal repeats the structure of the second signal level until the signal is interrupted.
\end{definition} 

We aim to improve the model's understanding of biological signaling pathways, enhancing its ability to learn and reason interactions within complex signal networks. We create prompts in the ProCoT (Protein Chain of Thought) format to incrementally decompose the signal transduction process like \autoref{def:signal}, simulating real pathways of signal propagation. Specifically, we clarify the structure knowledge within the signaling pathway, such as signaling proteins and effector proteins. We then design rules for signal transmission across different levels to simulate the iterative process of signal transduction in proteins. Finally, we also design signal interruptions to simulate the continuity of protein transmission in real life. We use hard code to convert protein interaction information as \autoref{def:procot} in dataset into ProCoT format natural language to imitate protein signaling pathways as \autoref{def:signal}. \autoref{fig:pcot_prompt} in the appendix is an example of ProCoT. The answer to this is "The relationship is activation". 

\subsubsection{Training on ProCot Format Dataset}
\begin{figure}
\centering
\vspace{-0.8cm}
\includegraphics[width=0.9\linewidth]{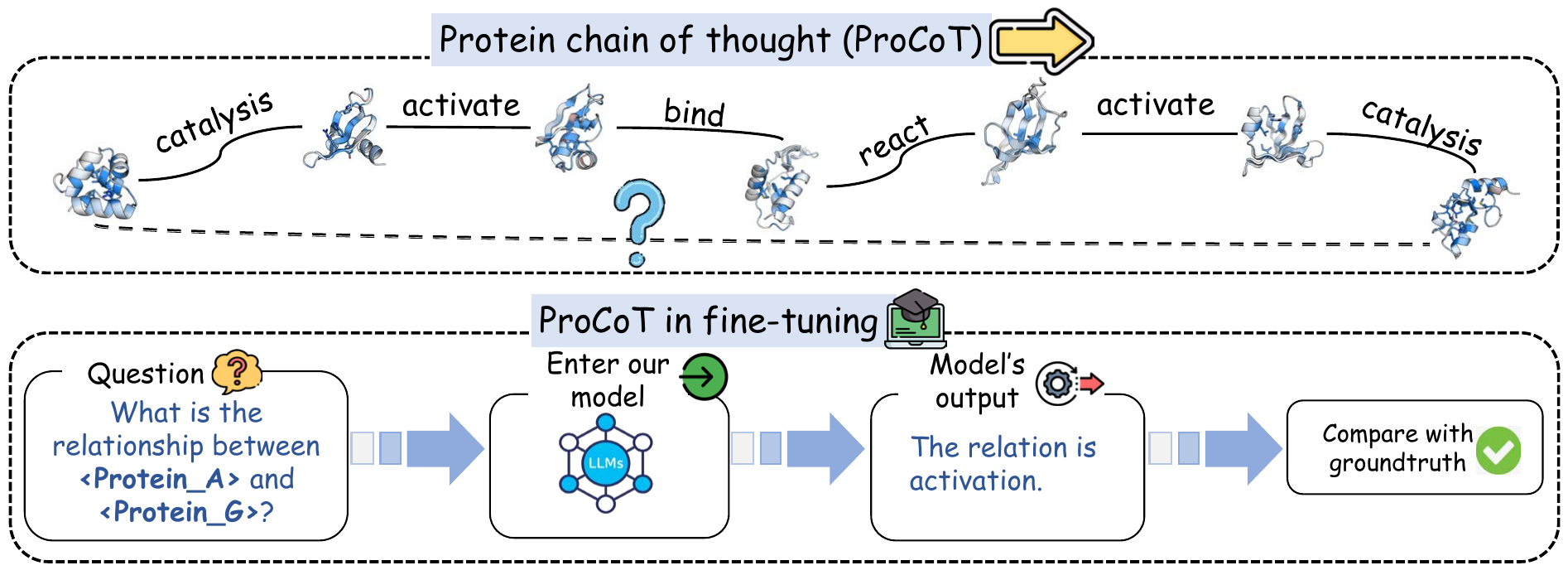}
\caption{The fine-tuning process of ProCoT. Within the first dashed box, solid lines between proteins represent the signaling pathway, and the dashed lines connecting the head and tail proteins indicate the masked interaction. Our model will predict the type of masked interaction.}
\label{fig:pcot_train}
\end{figure}

In biology, co-expression and co-localization refer to the phenomenon in which proteins that are often expressed or located together in the cell tend to participate in the same or interconnected biological processes~\citep{zhang2019method}. Thus, biologists frequently use known protein information to infer unknown protein interactions.
Based on the principles of co-expression and co-localization in biology, we have formulated our training strategy. After constructing our ProCoT training data using DFS, we will obtain a circular protein interaction
as in \autoref{fig:pcot_train}. We mask the relationship between the initial and final proteins in the signaling pathway and let our model predict this relationship. 

\subsubsection{Why ProCot Works: Biological Intuition behind ProCot}
In this section, we will explore ProCoT through the lenses of biology and AI.  \textbf{(1) Simulating Biological Signaling Pathways:} The biological signaling pathway is a series of ordered, rule-based interaction processes. During the process, the output of each step serves as the input for the next step, forming a highly organized information transmission chain. This pattern is consistent with the methods of handling serialized information in natural language processing (NLP). Additionally, Flan-T5's self-attention blocks can capture the indirect relationships between proteins that are distant and do not have direct interactions. It is crucial for understanding biological signal transduction processes that involve multiple steps and complex intermediate links. \textbf{(2) Signal Interruption Mechanism:} This mechanism aims to mimic protein adaptability since the signaling between proteins needs to be constantly interrupted to ensure the cell's accurate response to external changes. This mechanism we designed can satisfy the complex feedback mechanisms and regulatory networks within the cell. Overall, our ProCoT design follows biological principles.

\subsection{Enhancing Protein Sequence and Function Comprehension }
In our methodology, we enhance the model's understanding of protein sequence by replacing the T5 embedding with the ProTrans embedding vectors and also perform instruction fine-tuning to enable the model to learn protein functions. 

\subsubsection{Embedding replacement}
We use a new embedding mechanism facilitated by the \texttt{ProTrans}~\citep{elnaggar2021prottrans} model to perform embedding replacement. \texttt{ProTrans} 
 is a large-scale protein model capable of transforming protein sequences into embedding vectors with the biophysical and structural features of proteins.
\begin{definition}{(Embedding Replacement)}\label{def:emb}
Given the protein id \( P_{\text{id}} \), we can query its protein sequence \( P_{\text{seq}} \) by the professional bioinformatics tool Ensembl BioMart.  \( P_{\text{seq}} \) will be the input of ProTrans and ProTrans will output a $1 \times 1024$-dimensional vector \( P_{\text{emb}} \).
\end{definition}

We add all protein IDs in the protein dataset and their corresponding embeddings to the T5 vocabulary to solve the OOV (out of vocabulary) problem: The T5 vocabulary does not contain words for protein ID numbers. Because ProtTrans is trained based on T5-large, the process of adding to the vocabulary does not involve any change in dimensions, and there is no information loss. Additionally, this approach can also avoid the issue where a protein ID is tokenized into multiple tokens, causing the model to fail to understand it.

The replaced embedding vectors from ProtTrans can add a lot of prior knowledge about the intrinsic patterns and biophysical features of proteins to our model, better applying it to subsequent protein interaction prediction tasks.

\subsubsection{Instruction Finetuning}
Mol-Instructions (Mol)~\citep{fang2023mol} dataset is applied to conduct instruction fine-tuning on our model. Mol is a comprehensive instruction dataset designed for the field of biomolecules that aims to address the limitations of LLMs in biomolecular research. We use Mol to teach our model with knowledge of protein functions.

\begin{definition}{(Description of Mol)}
The content of the Mol dataset can be defined as follows: instructional text \( I \) related to LLMs queries, an input amino acid sequence \( S \) that includes essential protein information, and metadata \( M \) that sheds light on vital details like the protein's subcellular location, its primary function, and its participation in biological processes, followed by a corresponding output \( O \) that serves as the response to \( I \).
\end{definition}

To facilitate instruction-based fine-tuning, we convert these objects into prompt-answer pairs \((P, A)\). This is shown as \autoref{fig:instruction} in the appendix. We convert the entire Mol dataset into a prompt-answer format and using these prompt-answer pairs for instruction fine-tuning of our model(ProLLM). This part can enhance the model's predictive performance on protein-related tasks.

\section{Experiments}
In this section, we present the experimental results to answer the following research questions (\textbf{RQ}s).

\begin{itemize}[leftmargin=*]
\item \textbf{RQ1} - How does the performance of the proposed ProLLM framework compare to other baselines in terms of PPI prediction accuracy and generalizability?
\item \textbf{RQ2} - How do different LLM backbones (e.g., Flan-T5-base and LLaMA-7b) affect the performance of the proposed ProLLM framework?
\item \textbf{RQ3} - What are the contributions of each component to the PPI prediction performance of the ProLLM framework?
\end{itemize}

\subsection{Experimental Settings}
\paragraph{Datasets:}
We undertake comprehensive evaluations on a trio of public Protein-Protein Interaction (PPI) datasets: Human~\citep{song2022learning}, STRING~\citep{szklarczyk2019string}, SHS27k, and SHS148k~\citep{chen2019multifaceted}. In these four datasets, we employ DFS dataset partitioning techniques. By prioritizing depth, DFS can effectively capture the step-by-step signal transmission and the hierarchical signal level of protein signaling pathways. Each dataset is split into training, validation, and testing sets, maintaining a proportion of 70\%, 10\%, and 20\% respectively.

\paragraph{Baselines:}
Earlier studies on predicting PPI are not pre-trained, they do not have prior knowledge about the proteins. Hence, we choose SVM~\citep{romero2019ppi}, DPPI~\citep{hashemifar2018predicting}, DNN-PPI~\citep{li2018deep}, PIPR~\citep{chen2019multifaceted}, GNN-PPI~\citep{lv2021learning} and SemiGNN-PPI~\citep{zhao2023semignn} as the baseline without pre-training. Furthermore, we explore how protein pre-training can influence the PPI, we include ProBERT~\citep{elnaggar2021prottrans}, SM-1b~\citep{rives2021biological}, GearNet-Edge~\citep{zhang2022protein}, and KeAP~\citep{zhou2022protein} as pre-trained baselines. MAPE-PPI~\citep{wu2024mape} and InstrucGLM~\citep{ye2023language} have two versions: one with pre-training and one without pre-training.

\paragraph{Evaluation Metrics:}
We selected the micro-F1 score~\citep{harbecke2022only} as our evaluation metric because the PPI dataset exhibits class imbalance, making the F1 score a very relevant reference. Note that micro-F1 is widely used in the protein-protein interaction task~\citep{tran2018end}. Each dataset follows a 70\% training, 10\% validation, and 20\% testing data split. Subsequently, we will choose different random seeds for training and testing, conducting a total of 10 tests. The mean micro-F1 score across these trials will serve as the definitive measure of model performance, with the accompanying standard deviation reflecting variability in different experimental runs.

\subsection{Comparative Experiment (RQ1)}
We compare the performance of ProLLM with other baselines (w/
and w/o additional pre-training data) on four datasets in \autoref{tab:compare}. Note that here we use Flan-T5-large~\citep{T5} as the backbone model. Based on the results, we can make three important observations: (1) Our method outperforms other baselines without pre-training. Although InstructGLM is also LLM based, it lags behind ProLLM; (2) As for the models pre-trained on protein dataset, they cannot achieve the performance of ProLLM without prior knowledge; (3) ProLLM outperforms GearNet-Edge, KeAP, and MAPE-PPI, although they utilize a significantly larger dataset comprised of structural and knowledge graph data for pre-training than the Mol dataset. 

%\begin{table}[h!]
\begin{table}[]
\centering
\scriptsize
%\resizebox{\textwidth}{!}{
\vspace{-0.7cm}
\begin{tabular}{@{}c|c|cccc@{}}
\toprule
Method & Pre-training Dataset & \multicolumn{1}{c}{Human} & \multicolumn{1}{c}{SHS27k} & \multicolumn{1}{c}{SHS148k} & \multicolumn{1}{c}{STRING} \\ 
\midrule
SVM & - & 61.28$\pm$1.28 & 53.07$\pm$5.16 & 58.59$\pm$0.07 & 64.59$\pm$0.03 \\
DPPI & - & 54.19$\pm$0.78& 46.12$\pm$3.02 & 52.03$\pm$1.18& 66.82$\pm$0.29 \\
DNN-PPI & - & 61.72$\pm$1.30& 54.34$\pm$1.30  & 58.42$\pm$2.05 & 64.94$\pm$0.93 \\
PIPR & - & 62.72$\pm$0.50 & 57.80$\pm$3.24& 63.98$\pm$0.76 & 67.45$\pm$0.30\\
GNN-PPI & - & 78.61$\pm$1.38 & 66.52$\pm$5.26 & 75.34$\pm$1.54 & 84.28$\pm$0.89\\
SemiGNN-PPI & - & 80.79$\pm$1.40 & 69.25$\pm$3.91 &77.62$\pm$1.08 &  84.85$\pm$0.65 \\
MAPE-PPI & - & \underline{82.13$\pm$1.47} & \underline{72.04$\pm$3.46} &\underline{ 80.45$\pm$1.12} & \underline{ 86.48$\pm$0.52} \\
InstructGLM & - & 81.35$\pm$2.04& 70.01$\pm$3.75 & 75.35$\pm$1.98  & 84.15$\pm$1.85 \\
\textbf{ProLLM(Flan-T5-large)} & - & \textbf{87.32$\pm$1.93} & \textbf{75.13$\pm$3.76} & \textbf{85.13$\pm$1.86} & \textbf{87.12$\pm$1.68}\\
\midrule
ProBERT & BFD & 79.58$\pm$0.76 & 68.85$\pm$3.18& 74.76$\pm$1.21 & 83.82$\pm$0.49\\
EMS-1b & UniRef50 & 81.48$\pm$1.02& 70.69$\pm$3.40 & 79.64$\pm$1.93& 85.21$\pm$0.76 \\
KeAP & ProteinKG25 & 82.31$\pm$0.71& 72.38$\pm$2.96& 80.20$\pm$1.26 & 86.58$\pm$0.41\\
GearNet-Edge & AlphaFoldDB & 82.87$\pm$1.02& 72.06$\pm$3.56& 79.84$\pm$1.65 & 85.96$\pm$1.01\\
MAPE-PPI & CATH4.2 & 83.64$\pm$1.22 & 73.21$\pm$2.97&  81.78$\pm$1.24 &  \underline{87.23$\pm$0.35}  \\
InstructGLM & Mol dataset & \underline{85.71$\pm$2.01} & \underline{75.64$\pm$3.49 }& \underline{83.41$\pm$1.78} & 85.25$\pm$1.72 \\
\textbf{ProLLM(Flan-T5-large)} & Mol dataset & \textbf{91.05$\pm$1.63} & \textbf{78.09$\pm$3.24} & \textbf{87.66$\pm$1.68} & \textbf{89.21$\pm$1.45} \\
\bottomrule
\end{tabular}
%}
\caption{The micro-F1 of different methods (w/o and w/ additional pre-training data) on different datasets, where bold and underline denote the best and second best metrics, respectively. Higher micro-F1 denotes better performance.}
\label{tab:compare}
\end{table}

\subsection{Influence of Different Backbones (RQ2)}
We choose Flan-T5-base, Flan-T5-large, Flan-T5-XL~\citep{T5} and LLaMA-7b~\citep{touvron2023llama} models as the backbone for ProLLM. We report the micro-F1 performance comparison as in \autoref{tab:backbone}. We should replace the embedding in Flan-T5 with ProtTrans and ProtTrans is a pre-trained model based on Flan-T5-Large. Therefore, the embedding generated by ProtTrans will match Flan-T5-large better. Additionally, despite having more model parameters, LLaMA-v1-7b exhibits worse lower micro-F1 scores in four datasets compared to lighter models: Flan-T5-base, Flan-T5-large, and Flan-T5-XL. Furthermore, the greater standard deviation of LLaMA-v1-7b highlights its instability.

\begin{table}[]
\centering
\scriptsize
\begin{tabular}{@{}c|c|cccc@{}}
\toprule
Method & Pre-training Dataset & \multicolumn{1}{c}{Human} & \multicolumn{1}{c}{SHS27k} & \multicolumn{1}{c}{SHS148k} & \multicolumn{1}{c}{STRING} \\ 
\midrule
ProLLM-Flan-T5-base & - & 82.62$\pm$2.01 & 71.67$\pm$4.36 & 78.19$\pm$1.82 & 80.63$\pm$1.97 \\
\textbf{ProLLM-Flan-T5-large} & - & \textbf{87.32$\pm$1.93} & \textbf{75.13$\pm$3.76} & \textbf{85.13$\pm$1.86} & \textbf{87.12$\pm$1.68} \\
ProLLM-Flan-T5-XL & - & \underline{84.32$\pm$2.65} & \underline{73.28$\pm$3.96} & \underline{81.73$\pm$2.21} & \underline{82.94$\pm$1.93} \\
ProLLM-LLaMA-v1-7b & - & 81.75$\pm$4.46 & 70.33$\pm$6.52 & 77.08$\pm$3.59 & 79.41$\pm$2.87 \\
\midrule
ProLLM-Flan-T5-base & Mol dataset & 87.92$\pm$2.07 & 74.28$\pm$4.09 & 85.07$\pm$1.87 & 87.11$\pm$1.88 \\
\textbf{ProLLM-Flan-T5-large} & Mol dataset & \textbf{91.05$\pm$1.63} & \textbf{78.09$\pm$3.24} & \textbf{87.66$\pm$1.93} & \textbf{89.21$\pm$1.45} \\
ProLLM-Flan-T5-XL & Mol dataset & \underline{89.16$\pm$2.41} & \underline{75.96$\pm$3.38} & \underline{86.13$\pm$1.97} & \underline{87.97$\pm$1.78} \\
ProLLM-LLaMA-v1-7b & Mol dataset & 87.08$\pm$4.72 & 74.87$\pm$6.51 & 84.61$\pm$3.53 & 86.71$\pm$2.61 \\
\bottomrule
\end{tabular}
\caption{The micro-F1 score of ProLLM on different backbones. Higher micro-F1 denotes better performance. Bold and underline denote the best and second-best metrics, respectively.}
\label{tab:backbone}
\end{table}

\subsection{Ablation Study (RQ3)}
In our ablation study (\autoref{tab:ablation}), we evaluated the impact of different configurations on PPI prediction. The configurations are as follows: \textbf{ProLLM w/o ProCoT:} This setup shuffles ProCoT data, disrupting signaling pathways to mimic the model's performance without pathway understanding. It limits the model's ability to learn from signaling sequences. ProLLM relies on memorizing fixed relations instead of reasoning through intermediate protein relationships; \textbf{ProLLM w/o Embedding Replacement:} It will compare the model with replaced embeddings to the model without expended vocabulary, which evaluates the effect of protein-specific embedding features; \textbf{ProLLM w/o Instruction Fine-tuning:} This setup examines the model's capability to predict protein-protein interactions without the application of instruction fine-tuning on Mol dataset; \textbf{ProLLM w/o Embedding and Instruction Fine-tuning:} This configuration tests the model's performance without utilizing both protein-specific embedding features and instruction fine-tuning on Mol dataset. 

In our ablation study, we have identified that ProCoT has the most significant impact on the performance of ProLLM.  Our experiments revealed that introducing ProCoT led to substantial improvements in the performance of the model. Additionally, we explored other techniques such as embedding replacement and instruction fine-tuning on the Mol dataset. While these approaches did show some positive effects on the model's performance, their impact was found to be comparatively smaller when compared to the influence of ProCoT.

\begin{table}[]
\centering
\scriptsize
\begin{tabular}{@{}ccclcccc@{}}
\toprule
\multirow{2}{*}{ProCoT} & \multirow{2}{*}{\begin{tabular}[c]{@{}c@{}}Embedding\\ Replacement\end{tabular}} & \multirow{2}{*}{\begin{tabular}[c]{@{}c@{}}Instruction \\ Fine-tuning\end{tabular}} &  & \multicolumn{4}{c}{Dataset}       \\ \cmidrule(l){5-8} 
                        &                                                                                  &                                                                                     &  & Human & SHS27k & SHS148k & STRING \\ \midrule
\checkmark              &                                                                                  &                                                                                     &  & 83.87$\pm$1.25  & 70.83$\pm$1.54  & 79.12$\pm$2.64   & 82.68$\pm$1.36  \\
\checkmark              & \checkmark                                                                      &                                                                                     &  & 87.32$\pm$1.93 & 74.26$\pm$3.53  & 85.13$\pm$1.86   & 87.12$\pm$1.68  \\
\checkmark              &                                                                                  & \checkmark                                                                           &  & 88.53$\pm$1.45 & 73.59$\pm$2.29  & 85.94$\pm$3.39   & 87.62$\pm$1.68  \\
                        & \checkmark                                                                      & \checkmark                                                                           &  & 78.32$\pm$2.65 & 61.98$\pm$2.06  & 74.10$\pm$1.27   & 77.85$\pm$1.54  \\
\checkmark              & \checkmark                                                                      & \checkmark                                                                           &  & \textbf{91.03$\pm$1.63} & \textbf{78.09$\pm$3.24}  & \textbf{87.64$\pm$1.93}   & \textbf{89.28$\pm$1.45}  \\ \bottomrule
\end{tabular}
\caption{Ablation study. The metric here is micro-F1. Where bold denote
the best metrics. Higher micro-F1 denotes better performance.}
\label{tab:ablation}
\end{table}

\section{Conclusions and Future Work}
In this paper, we propose ProLLM, a novel framework that leverages LLMs for protein-protein interaction prediction by representing protein data in natural language formats. Our key contributions include: 1) ProCoT (Protein Chain of Thought) to convert multi-step protein signaling pathways to natural language prompts, and the design of ProCoT can reflect the actual protein signaling passing within a biological organism. Additionally, the format of ProCoT is sequential, which is a type of information that LLMs are good at processing.  2) Integration of protein-specific embeddings from ProtTrans, and 3) Instruction fine-tuning on protein knowledge datasets. Through extensive experiments on four PPI datasets, ProLLM significantly outperformed existing graph-based and language model methods in prediction accuracy and generalizability. By unifying language models with structured biological data, our work opens up new possibilities for driving discoveries in computational biology, drug discovery, and broader scientific domains.

\subsection{Acknowledgement}
We thank Wenyue Hua, Kai Mei and Taowen Wang for their valuable discussions and suggestions
during the project.

\bibliography{colm2024_conference}

\begin{thebibliography}{56}
\providecommand{\natexlab}[1]{#1}
\providecommand{\url}[1]{\texttt{#1}}
\expandafter\ifx\csname urlstyle\endcsname\relax
  \providecommand{\doi}[1]{doi: #1}\else
  \providecommand{\doi}{doi: \begingroup \urlstyle{rm}\Url}\fi

\bibitem[Br{\"u}ckner et~al.(2009)Br{\"u}ckner, Polge, Lentze, Auerbach, and Schlattner]{bruckner2009yeast}
Anna Br{\"u}ckner, C{\'e}cile Polge, Nicolas Lentze, Daniel Auerbach, and Uwe Schlattner.
\newblock Yeast two-hybrid, a powerful tool for systems biology.
\newblock \emph{International journal of molecular sciences}, 10\penalty0 (6):\penalty0 2763--2788, 2009.

\bibitem[Chen et~al.(2019)Chen, Ju, Zhou, Chen, Zhang, Chang, Zaniolo, and Wang]{chen2019multifaceted}
Muhao Chen, Chelsea J-T Ju, Guangyu Zhou, Xuelu Chen, Tianran Zhang, Kai-Wei Chang, Carlo Zaniolo, and Wei Wang.
\newblock Multifaceted protein--protein interaction prediction based on siamese residual rcnn.
\newblock \emph{Bioinformatics}, 35\penalty0 (14):\penalty0 i305--i314, 2019.

\bibitem[Devlin et~al.(2018)Devlin, Chang, Lee, and Toutanova]{BERT}
Jacob Devlin, Ming-Wei Chang, Kenton Lee, and Kristina Toutanova.
\newblock Bert: Pre-training of deep bidirectional transformers for language understanding.
\newblock \emph{arXiv preprint arXiv:1810.04805}, 2018.

\bibitem[Du et~al.(2017)Du, Sun, Hu, Yao, Yan, and Zhang]{du2017deepppi}
Xiuquan Du, Shiwei Sun, Changlin Hu, Yu~Yao, Yuanting Yan, and Yanping Zhang.
\newblock Deepppi: boosting prediction of protein--protein interactions with deep neural networks.
\newblock \emph{Journal of chemical information and modeling}, 57\penalty0 (6):\penalty0 1499--1510, 2017.

\bibitem[Elnaggar et~al.(2021{\natexlab{a}})Elnaggar, Heinzinger, Dallago, Rehawi, Wang, Jones, Gibbs, Feher, Angerer, Steinegger, et~al.]{elnaggar2021prottrans}
Ahmed Elnaggar, Michael Heinzinger, Christian Dallago, Ghalia Rehawi, Yu~Wang, Llion Jones, Tom Gibbs, Tamas Feher, Christoph Angerer, Martin Steinegger, et~al.
\newblock Prottrans: Toward understanding the language of life through self-supervised learning.
\newblock \emph{IEEE transactions on pattern analysis and machine intelligence}, 44\penalty0 (10):\penalty0 7112--7127, 2021{\natexlab{a}}.

\bibitem[Elnaggar et~al.(2021{\natexlab{b}})Elnaggar, Heinzinger, Dallago, Rehawi, Yu, Jones, Gibbs, Feher, Angerer, Steinegger, Bhowmik, and Rost]{9477085}
Ahmed Elnaggar, Michael Heinzinger, Christian Dallago, Ghalia Rehawi, Wang Yu, Llion Jones, Tom Gibbs, Tamas Feher, Christoph Angerer, Martin Steinegger, Debsindhu Bhowmik, and Burkhard Rost.
\newblock Prottrans: Towards cracking the language of lifes code through self-supervised deep learning and high performance computing.
\newblock \emph{IEEE Transactions on Pattern Analysis and Machine Intelligence}, pp.\  1--1, 2021{\natexlab{b}}.
\newblock \doi{10.1109/TPAMI.2021.3095381}.

\bibitem[Fan et~al.(2024)Fan, Hua, Li, Zhu, Jin, Li, Ling, Chi, Wang, Ma, et~al.]{fan2024nphardeval4v}
Lizhou Fan, Wenyue Hua, Xiang Li, Kaijie Zhu, Mingyu Jin, Lingyao Li, Haoyang Ling, Jinkui Chi, Jindong Wang, Xin Ma, et~al.
\newblock Nphardeval4v: A dynamic reasoning benchmark of multimodal large language models.
\newblock \emph{arXiv preprint arXiv:2403.01777}, 2024.

\bibitem[Fang et~al.(2023)Fang, Liang, Zhang, Liu, Huang, Chen, Fan, and Chen]{fang2023mol}
Yin Fang, Xiaozhuan Liang, Ningyu Zhang, Kangwei Liu, Rui Huang, Zhuo Chen, Xiaohui Fan, and Huajun Chen.
\newblock Mol-instructions: A large-scale biomolecular instruction dataset for large language models.
\newblock \emph{arXiv preprint arXiv:2306.08018}, 2023.

\bibitem[Gao et~al.(2023)Gao, Jiang, Zhang, Jiang, Li, Zhao, Yang, Huang, and Li]{gao2023hierarchical}
Ziqi Gao, Chenran Jiang, Jiawen Zhang, Xiaosen Jiang, Lanqing Li, Peilin Zhao, Huanming Yang, Yong Huang, and Jia Li.
\newblock Hierarchical graph learning for protein--protein interaction.
\newblock \emph{Nature Communications}, 14\penalty0 (1):\penalty0 1093, 2023.

\bibitem[Guo et~al.(2008)Guo, Yu, Wen, and Li]{guo2008using}
Yanzhi Guo, Lezheng Yu, Zhining Wen, and Menglong Li.
\newblock Using support vector machine combined with auto covariance to predict protein--protein interactions from protein sequences.
\newblock \emph{Nucleic acids research}, 36\penalty0 (9):\penalty0 3025--3030, 2008.

\bibitem[Harbecke et~al.(2022)Harbecke, Chen, Hennig, and Alt]{harbecke2022only}
David Harbecke, Yuxuan Chen, Leonhard Hennig, and Christoph Alt.
\newblock Why only micro-f1? class weighting of measures for relation classification.
\newblock \emph{arXiv preprint arXiv:2205.09460}, 2022.

\bibitem[Hashemifar et~al.(2018)Hashemifar, Neyshabur, Khan, and Xu]{hashemifar2018predicting}
Somaye Hashemifar, Behnam Neyshabur, Aly~A Khan, and Jinbo Xu.
\newblock Predicting protein--protein interactions through sequence-based deep learning.
\newblock \emph{Bioinformatics}, 34\penalty0 (17):\penalty0 i802--i810, 2018.

\bibitem[Hou et~al.(2023)Hou, Wang, Bu, Wang, and Sun]{hou2023emngly}
Xiaoyang Hou, Yu~Wang, Dongbo Bu, Yaojun Wang, and Shiwei Sun.
\newblock Emngly: predicting n-linked glycosylation sites using the language models for feature extraction.
\newblock \emph{Bioinformatics}, 39\penalty0 (11):\penalty0 btad650, 2023.

\bibitem[Hsu et~al.(2022)Hsu, Verkuil, Liu, Lin, Hie, Sercu, Lerer, and Rives]{hsu2022learning}
Chloe Hsu, Robert Verkuil, Jason Liu, Zeming Lin, Brian Hie, Tom Sercu, Adam Lerer, and Alexander Rives.
\newblock Learning inverse folding from millions of predicted structures.
\newblock \emph{ICML}, 2022.
\newblock \doi{10.1101/2022.04.10.487779}.
\newblock URL \url{https://www.biorxiv.org/content/early/2022/04/10/2022.04.10.487779}.

\bibitem[Hua et~al.(2024)Hua, Zhu, Li, Fan, Lin, Jin, Xue, Li, Wang, and Zhang]{hua2024disentangling}
Wenyue Hua, Kaijie Zhu, Lingyao Li, Lizhou Fan, Shuhang Lin, Mingyu Jin, Haochen Xue, Zelong Li, JinDong Wang, and Yongfeng Zhang.
\newblock Disentangling logic: The role of context in large language model reasoning capabilities.
\newblock \emph{arXiv preprint arXiv:2406.02787}, 2024.

\bibitem[Ito et~al.(2001)Ito, Chiba, Ozawa, Yoshida, Hattori, and Sakaki]{ito2001comprehensive}
Takashi Ito, Tomoko Chiba, Ritsuko Ozawa, Mikio Yoshida, Masahira Hattori, and Yoshiyuki Sakaki.
\newblock A comprehensive two-hybrid analysis to explore the yeast protein interactome.
\newblock \emph{Proceedings of the National Academy of Sciences}, 98\penalty0 (8):\penalty0 4569--4574, 2001.

\bibitem[Jin et~al.(2024{\natexlab{a}})Jin, Yu, Zhang, Shu, Zhu, Du, Zhang, and Meng]{jin2024health}
Mingyu Jin, Qinkai Yu, Chong Zhang, Dong Shu, Suiyuan Zhu, Mengnan Du, Yongfeng Zhang, and Yanda Meng.
\newblock Health-llm: Personalized retrieval-augmented disease prediction model.
\newblock \emph{arXiv preprint arXiv:2402.00746}, 2024{\natexlab{a}}.

\bibitem[Jin et~al.(2024{\natexlab{b}})Jin, Yu, Zhao, Hua, Meng, Zhang, Du, et~al.]{jin2024impact}
Mingyu Jin, Qinkai Yu, Haiyan Zhao, Wenyue Hua, Yanda Meng, Yongfeng Zhang, Mengnan Du, et~al.
\newblock The impact of reasoning step length on large language models.
\newblock \emph{arXiv preprint arXiv:2401.04925}, 2024{\natexlab{b}}.

\bibitem[Li et~al.(2018)Li, Gong, Yu, and Zhou]{li2018deep}
Hang Li, Xiu-Jun Gong, Hua Yu, and Chang Zhou.
\newblock Deep neural network based predictions of protein interactions using primary sequences.
\newblock \emph{Molecules}, 23\penalty0 (8):\penalty0 1923, 2018.

\bibitem[Lin et~al.(2023)Lin, Akin, Rao, Hie, Zhu, Lu, Smetanin, Verkuil, Kabeli, Shmueli, et~al.]{lin2023evolutionary}
Zeming Lin, Halil Akin, Roshan Rao, Brian Hie, Zhongkai Zhu, Wenting Lu, Nikita Smetanin, Robert Verkuil, Ori Kabeli, Yaniv Shmueli, et~al.
\newblock Evolutionary-scale prediction of atomic-level protein structure with a language model.
\newblock \emph{Science}, 379\penalty0 (6637):\penalty0 1123--1130, 2023.

\bibitem[Liu et~al.(2021)Liu, Luo, Li, Song, and Peng]{liu2021deep}
Xianggen Liu, Yunan Luo, Pengyong Li, Sen Song, and Jian Peng.
\newblock Deep geometric representations for modeling effects of mutations on protein-protein binding affinity.
\newblock \emph{PLoS computational biology}, 17\penalty0 (8):\penalty0 e1009284, 2021.

\bibitem[Long et~al.(2022)Long, Wu, Liu, Fang, Kwoh, Chen, Luo, and Li]{long2022pre}
Yahui Long, Min Wu, Yong Liu, Yuan Fang, Chee~Keong Kwoh, Jinmiao Chen, Jiawei Luo, and Xiaoli Li.
\newblock Pre-training graph neural networks for link prediction in biomedical networks.
\newblock \emph{Bioinformatics}, 38\penalty0 (8):\penalty0 2254--2262, 2022.

\bibitem[Lv et~al.(2021)Lv, Hu, Bi, and Zhang]{lv2021learning}
Guofeng Lv, Zhiqiang Hu, Yanguang Bi, and Shaoting Zhang.
\newblock Learning unknown from correlations: graph neural network for inter-novel-protein interaction prediction.
\newblock \emph{arXiv preprint arXiv:2105.06709}, 2021.

\bibitem[Mann et~al.(2001)Mann, Hendrickson, and Pandey]{mann2001analysis}
Matthias Mann, Ronald~C Hendrickson, and Akhilesh Pandey.
\newblock Analysis of proteins and proteomes by mass spectrometry.
\newblock \emph{Annual review of biochemistry}, 70\penalty0 (1):\penalty0 437--473, 2001.

\bibitem[O'Neil et~al.(2017)O'Neil, Bailey, and Hieter]{o2017synthetic}
Nigel~J O'Neil, Melanie~L Bailey, and Philip Hieter.
\newblock Synthetic lethality and cancer.
\newblock \emph{Nature Reviews Genetics}, 18\penalty0 (10):\penalty0 613--623, 2017.

\bibitem[Peng et~al.(2023)Peng, Li, He, Galley, and Gao]{gpt4}
Baolin Peng, Chunyuan Li, Pengcheng He, Michel Galley, and Jianfeng Gao.
\newblock Instruction tuning with gpt-4.
\newblock \emph{arXiv preprint arXiv:2304.03277}, 2023.

\bibitem[Raffel et~al.(2020)Raffel, Shazeer, Roberts, Lee, Narang, Matena, Zhou, Li, and Liu]{T5}
Colin Raffel, Noam Shazeer, Adam Roberts, Katherine Lee, Sharan Narang, Michael Matena, Yanqi Zhou, Wei Li, and Peter~J Liu.
\newblock Exploring the limits of transfer learning with a unified text-to-text transformer.
\newblock \emph{The Journal of Machine Learning Research}, 21\penalty0 (1):\penalty0 5485--5551, 2020.

\bibitem[R{\'e}au et~al.(2023)R{\'e}au, Renaud, Xue, and Bonvin]{reau2023deeprank}
Manon R{\'e}au, Nicolas Renaud, Li~C Xue, and Alexandre~MJJ Bonvin.
\newblock Deeprank-gnn: a graph neural network framework to learn patterns in protein--protein interfaces.
\newblock \emph{Bioinformatics}, 39\penalty0 (1):\penalty0 btac759, 2023.

\bibitem[Rives et~al.(2021)Rives, Meier, Sercu, Goyal, Lin, Liu, Guo, Ott, Zitnick, Ma, et~al.]{rives2021biological}
Alexander Rives, Joshua Meier, Tom Sercu, Siddharth Goyal, Zeming Lin, Jason Liu, Demi Guo, Myle Ott, C~Lawrence Zitnick, Jerry Ma, et~al.
\newblock Biological structure and function emerge from scaling unsupervised learning to 250 million protein sequences.
\newblock \emph{Proceedings of the National Academy of Sciences}, 118\penalty0 (15):\penalty0 e2016239118, 2021.

\bibitem[Rodrigues et~al.(2021)Rodrigues, Pires, and Ascher]{rodrigues2021mmcsm}
Carlos~HM Rodrigues, Douglas~EV Pires, and David~B Ascher.
\newblock mmcsm-ppi: predicting the effects of multiple point mutations on protein--protein interactions.
\newblock \emph{Nucleic Acids Research}, 49\penalty0 (W1):\penalty0 W417--W424, 2021.

\bibitem[Romero-Molina et~al.(2019)Romero-Molina, Ruiz-Blanco, Harms, M{\"u}nch, and Sanchez-Garcia]{romero2019ppi}
Sandra Romero-Molina, Yasser~B Ruiz-Blanco, Mirja Harms, Jan M{\"u}nch, and Elsa Sanchez-Garcia.
\newblock Ppi-detect: A support vector machine model for sequence-based prediction of protein--protein interactions.
\newblock \emph{Journal of computational chemistry}, 40\penalty0 (11):\penalty0 1233--1242, 2019.

\bibitem[Rotilio et~al.(2012)Rotilio, Della~Corte, D'Imperio, Coletta, Marcone, Silvestri, Giordano, Di~Michele, and Donati]{rotilio2012proteomics}
Domenico Rotilio, Anna Della~Corte, Marco D'Imperio, Walter Coletta, Simone Marcone, Cristian Silvestri, Lucia Giordano, Michela Di~Michele, and Maria~Benedetta Donati.
\newblock Proteomics: bases for protein complexity understanding.
\newblock \emph{Thrombosis research}, 129\penalty0 (3):\penalty0 257--262, 2012.

\bibitem[Shen et~al.(2007)Shen, Zhang, Luo, Zhu, Yu, Chen, Li, and Jiang]{shen2007predicting}
Juwen Shen, Jian Zhang, Xiaomin Luo, Weiliang Zhu, Kunqian Yu, Kaixian Chen, Yixue Li, and Hualiang Jiang.
\newblock Predicting protein--protein interactions based only on sequences information.
\newblock \emph{Proceedings of the National Academy of Sciences}, 104\penalty0 (11):\penalty0 4337--4341, 2007.

\bibitem[Shengyuan et~al.(2024)Shengyuan, Cai, Fang, Huang, and Sun]{shengyuan2024differentiable}
Chen Shengyuan, Yunfeng Cai, Huang Fang, Xiao Huang, and Mingming Sun.
\newblock Differentiable neuro-symbolic reasoning on large-scale knowledge graphs.
\newblock \emph{Advances in Neural Information Processing Systems}, 36, 2024.

\bibitem[Shu et~al.(2024)Shu, Chen, Jin, Zhang, Du, and Zhang]{shu2024knowledge}
Dong Shu, Tianle Chen, Mingyu Jin, Yiting Zhang, Mengnan Du, and Yongfeng Zhang.
\newblock Knowledge graph large language model (kg-llm) for link prediction, 2024.

\bibitem[Song et~al.(2022)Song, Luo, Luo, Liu, Niu, and Zeng]{song2022learning}
Bosheng Song, Xiaoyan Luo, Xiaoli Luo, Yuansheng Liu, Zhangming Niu, and Xiangxiang Zeng.
\newblock Learning spatial structures of proteins improves protein--protein interaction prediction.
\newblock \emph{Briefings in bioinformatics}, 23\penalty0 (2):\penalty0 bbab558, 2022.

\bibitem[Sun et~al.(2017)Sun, Zhou, Lai, and Pei]{sun2017sequence}
Tanlin Sun, Bo~Zhou, Luhua Lai, and Jianfeng Pei.
\newblock Sequence-based prediction of protein protein interaction using a deep-learning algorithm.
\newblock \emph{BMC bioinformatics}, 18:\penalty0 1--8, 2017.

\bibitem[Szklarczyk et~al.(2019)Szklarczyk, Gable, Lyon, Junge, Wyder, Huerta-Cepas, Simonovic, Doncheva, Morris, Bork, et~al.]{szklarczyk2019string}
Damian Szklarczyk, Annika~L Gable, David Lyon, Alexander Junge, Stefan Wyder, Jaime Huerta-Cepas, Milan Simonovic, Nadezhda~T Doncheva, John~H Morris, Peer Bork, et~al.
\newblock String v11: protein--protein association networks with increased coverage, supporting functional discovery in genome-wide experimental datasets.
\newblock \emph{Nucleic acids research}, 47\penalty0 (D1):\penalty0 D607--D613, 2019.

\bibitem[Teufel et~al.(2022)Teufel, Almagro~Armenteros, Johansen, G{\'\i}slason, Pihl, Tsirigos, Winther, Brunak, von Heijne, and Nielsen]{teufel2022signalp}
Felix Teufel, Jos{\'e}~Juan Almagro~Armenteros, Alexander~Rosenberg Johansen, Magn{\'u}s~Halld{\'o}r G{\'\i}slason, Silas~Irby Pihl, Konstantinos~D Tsirigos, Ole Winther, S{\o}ren Brunak, Gunnar von Heijne, and Henrik Nielsen.
\newblock Signalp 6.0 predicts all five types of signal peptides using protein language models.
\newblock \emph{Nature biotechnology}, 40\penalty0 (7):\penalty0 1023--1025, 2022.

\bibitem[Thumuluri et~al.(2022)Thumuluri, Almagro~Armenteros, Johansen, Nielsen, and Winther]{thumuluri2022deeploc}
Vineet Thumuluri, Jos{\'e}~Juan Almagro~Armenteros, Alexander~Rosenberg Johansen, Henrik Nielsen, and Ole Winther.
\newblock Deeploc 2.0: multi-label subcellular localization prediction using protein language models.
\newblock \emph{Nucleic Acids Research}, 50\penalty0 (W1):\penalty0 W228--W234, 2022.

\bibitem[Touvron et~al.(2023)Touvron, Lavril, Izacard, Martinet, Lachaux, Lacroix, Rozi{\`e}re, Goyal, Hambro, Azhar, et~al.]{touvron2023llama}
Hugo Touvron, Thibaut Lavril, Gautier Izacard, Xavier Martinet, Marie-Anne Lachaux, Timoth{\'e}e Lacroix, Baptiste Rozi{\`e}re, Naman Goyal, Eric Hambro, Faisal Azhar, et~al.
\newblock Llama: Open and efficient foundation language models.
\newblock \emph{arXiv preprint arXiv:2302.13971}, 2023.

\bibitem[Tran \& Kavuluru(2018)Tran and Kavuluru]{tran2018end}
Tung Tran and Ramakanth Kavuluru.
\newblock An end-to-end deep learning architecture for extracting protein--protein interactions affected by genetic mutations.
\newblock \emph{Database}, 2018:\penalty0 1--13, 2018.

\bibitem[Wei et~al.(2022)Wei, Tay, Bommasani, Raffel, Zoph, Borgeaud, Yogatama, Bosma, Zhou, Metzler, et~al.]{wei2022emergent}
Jason Wei, Yi~Tay, Rishi Bommasani, Colin Raffel, Barret Zoph, Sebastian Borgeaud, Dani Yogatama, Maarten Bosma, Denny Zhou, Donald Metzler, et~al.
\newblock Emergent abilities of large language models.
\newblock \emph{arXiv preprint arXiv:2206.07682}, 2022.

\bibitem[Wilm(2009)]{wilm2009quantitative}
Matthias Wilm.
\newblock Quantitative proteomics in biological research.
\newblock \emph{Proteomics}, 9\penalty0 (20):\penalty0 4590--4605, 2009.

\bibitem[Wu et~al.(2024)Wu, Tian, Huang, Li, Lin, Chawla, and Li]{wu2024mape}
Lirong Wu, Yijun Tian, Yufei Huang, Siyuan Li, Haitao Lin, Nitesh~V Chawla, and Stan~Z Li.
\newblock Mape-ppi: Towards effective and efficient protein-protein interaction prediction via microenvironment-aware protein embedding.
\newblock In \emph{The Twelfth International Conference on Learning Representations}, 2024.
\newblock URL \url{https://openreview.net/forum?id=itGkF993gz}.

\bibitem[Xiao et~al.(2021)Xiao, Qiu, Li, Hsieh, and Tang]{xiao2021modeling}
Yijia Xiao, Jiezhong Qiu, Ziang Li, Chang-Yu Hsieh, and Jie Tang.
\newblock Modeling protein using large-scale pretrain language model, 2021.

\bibitem[Yang et~al.(2023)Yang, Jin, Tang, Han, Feng, Jiang, Zhong, Yin, and Hu]{yang2023harnessing}
Jingfeng Yang, Hongye Jin, Ruixiang Tang, Xiaotian Han, Qizhang Feng, Haoming Jiang, Shaochen Zhong, Bing Yin, and Xia Hu.
\newblock Harnessing the power of llms in practice: A survey on chatgpt and beyond.
\newblock \emph{ACM Transactions on Knowledge Discovery from Data}, 2023.

\bibitem[Ye et~al.(2024)Ye, Zhang, Wang, Xu, and Zhang]{ye2023language}
Ruosong Ye, Caiqi Zhang, Runhui Wang, Shuyuan Xu, and Yongfeng Zhang.
\newblock Language is all a graph needs.
\newblock \emph{EACL}, 2024.

\bibitem[You et~al.(2013)You, Lei, Zhu, Xia, and Wang]{you2013prediction}
Zhu-Hong You, Ying-Ke Lei, Lin Zhu, Junfeng Xia, and Bing Wang.
\newblock Prediction of protein-protein interactions from amino acid sequences with ensemble extreme learning machines and principal component analysis.
\newblock In \emph{BMC bioinformatics}, volume~14, pp.\  1--11. Springer, 2013.

\bibitem[Zhang et~al.(2019)Zhang, Zhong, Huang, Lin, and Wang]{zhang2019method}
Jinxiong Zhang, Cheng Zhong, Yiran Huang, Hai~Xiang Lin, and Mian Wang.
\newblock A method for identifying protein complexes with the features of joint co-localization and joint co-expression in static ppi networks.
\newblock \emph{Computers in Biology and Medicine}, 111:\penalty0 103333, 2019.

\bibitem[Zhang et~al.(2022)Zhang, Xu, Jamasb, Chenthamarakshan, Lozano, Das, and Tang]{zhang2022protein}
Zuobai Zhang, Minghao Xu, Arian Jamasb, Vijil Chenthamarakshan, Aurelie Lozano, Payel Das, and Jian Tang.
\newblock Protein representation learning by geometric structure pretraining.
\newblock \emph{arXiv preprint arXiv:2203.06125}, 2022.

\bibitem[Zhao et~al.(2023)Zhao, Qian, Yang, Zeng, Guan, Tam, and Li]{zhao2023semignn}
Ziyuan Zhao, Peisheng Qian, Xulei Yang, Zeng Zeng, Cuntai Guan, Wai~Leong Tam, and Xiaoli Li.
\newblock Semignn-ppi: Self-ensembling multi-graph neural network for efficient and generalizable protein-protein interaction prediction.
\newblock \emph{arXiv preprint arXiv:2305.08316}, 2023.

\bibitem[Zhou et~al.(2022)Zhou, Fu, Zhang, Cheng, and Yu]{zhou2022protein}
Hong-Yu Zhou, Yunxiang Fu, Zhicheng Zhang, Bian Cheng, and Yizhou Yu.
\newblock Protein representation learning via knowledge enhanced primary structure reasoning.
\newblock In \emph{The Eleventh International Conference on Learning Representations}, 2022.

\bibitem[Zhou et~al.(2023)Zhou, Fu, Zhang, Bian, and Yu]{zhou2023protein}
Hong-Yu Zhou, Yunxiang Fu, Zhicheng Zhang, Cheng Bian, and Yizhou Yu.
\newblock Protein representation learning via knowledge enhanced primary structure modeling, 2023.

\bibitem[Zhou et~al.(2020)Zhou, Cui, Hu, Zhang, Yang, Liu, Wang, Li, and Sun]{zhou2020graph}
Jie Zhou, Ganqu Cui, Shengding Hu, Zhengyan Zhang, Cheng Yang, Zhiyuan Liu, Lifeng Wang, Changcheng Li, and Maosong Sun.
\newblock Graph neural networks: A review of methods and applications.
\newblock \emph{AI open}, 1:\penalty0 57--81, 2020.

\bibitem[Zhuo et~al.(2024)Zhuo, Chi, Xu, Huang, Zheng, He, Mao, and Zhang]{zhuo2024protllm}
Le~Zhuo, Zewen Chi, Minghao Xu, Heyan Huang, Heqi Zheng, Conghui He, Xian-Ling Mao, and Wentao Zhang.
\newblock Protllm: An interleaved protein-language llm with protein-as-word pre-training.
\newblock \emph{arXiv preprint arXiv:2403.07920}, 2024.

\end{thebibliography}
\bibliographystyle{colm2024_conference}
\newpage
\section{Appendix}
\subsection{Dataset Partition Algorithm}
\begin{figure}[H] % 使用 [H] 来固定位置
\centering
\includegraphics[width=0.5\linewidth]{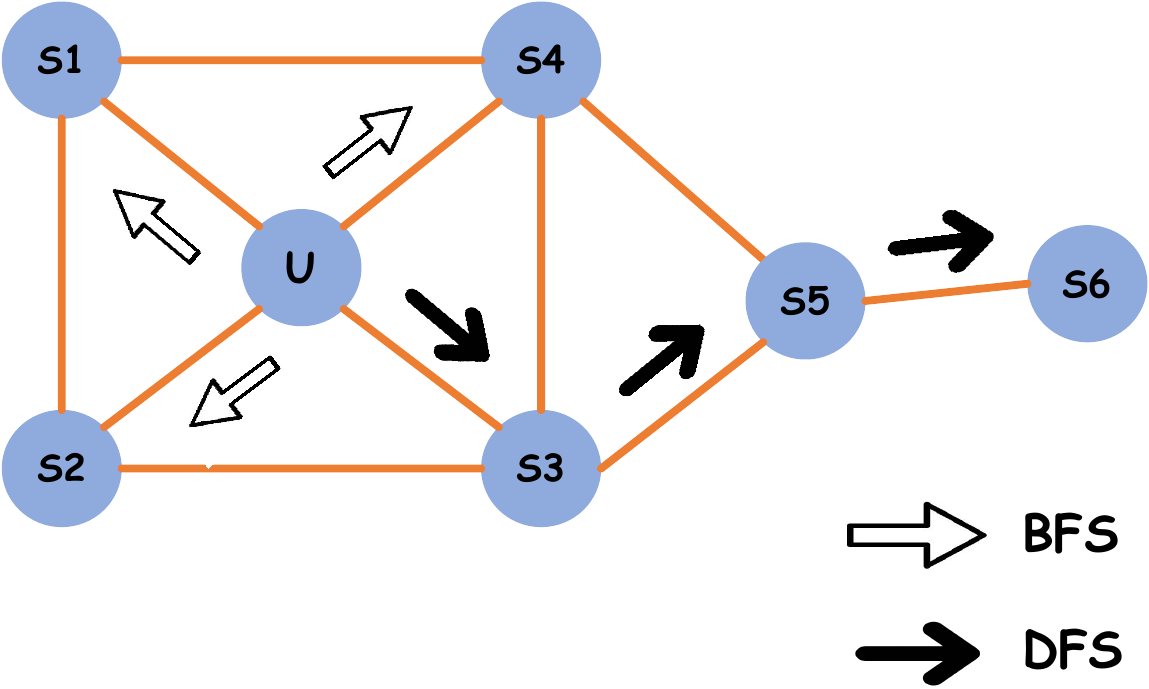}
\caption{Demo of BFS and DFS dataset partition method.}
\label{fig:bfsdfs}
\end{figure}

In the field of graph learning, DFS (Depth-First Search), BFS (Breadth-First Search), and random sampling are three common graph traversal or sampling strategies. \autoref{fig:bfsdfs} shows the difference in between DFS and BFS. 

\textbf{Depth-First Search (DFS):} DFS starts from a starting node and explores the graph's depth until it cannot go further, then backtracks to the nearest unvisited node from the starting node. This approach makes DFS inclined to explore the deep structure of the graph.
\textbf{Breadth-First Search (BFS):} BFS starts from a starting node and visits its neighboring nodes one by one, then proceeds to visit the neighboring nodes' neighboring nodes, and so on. This approach prioritizes exploring the breadth of the graph.
\textbf{Random Sampling:} Random sampling is a method of randomly selecting nodes for traversal or sampling. It can employ uniform random selection or select nodes based on certain probabilities.

The choice of strategy depends on the specific problem requirements and DFS is for exploring entire connected components. DFS can be used to find paths in a graph, especially when finding all paths from one node to another. DFS selects the next node for in-depth exploration at each step until the target node is found or cannot continue deeper. This is very similar to the protein signaling pathway in biology. From one protein to the target protein through different proteins, to simulate the signaling pathway. Additionally, after the DFS, we can obtain a cyclic structure of connected proteins, where one side of the cycle represents the signaling pathway between all proteins from the head protein to the tail protein, and the other side represents the direct interaction between the head and tail proteins. This is the data format we need in training our model. To simulate signaling pathways for training, we propose ProCot, and we \textbf{only use DFS} for dataset partition.

\subsection{The disscusion about the embedding replacement}
\begin{figure}[H] % 使用 [H] 来固定位置
\centering
\includegraphics[width=0.55\linewidth]{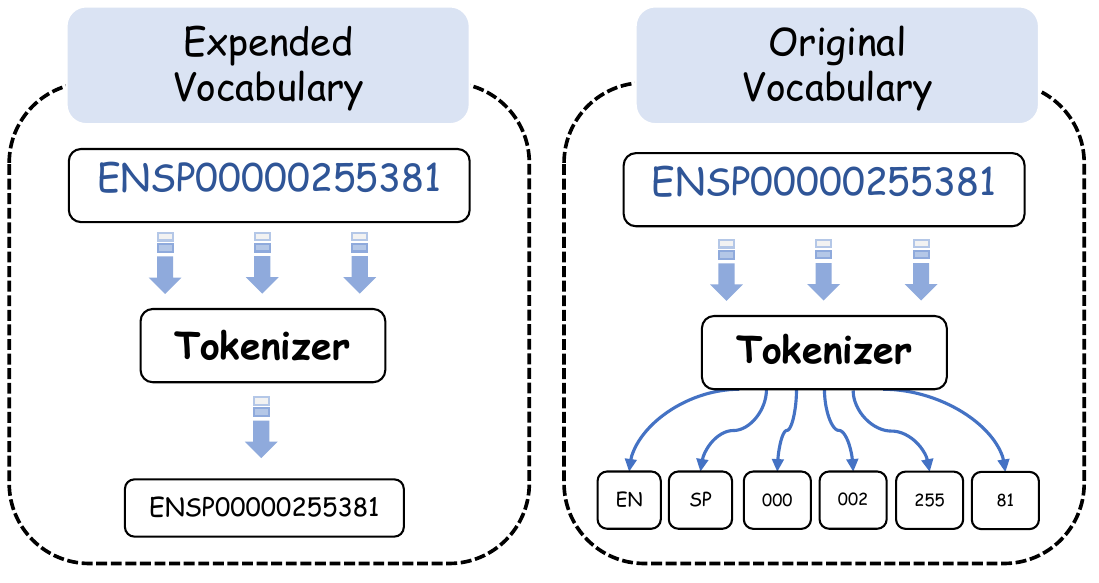}
\caption{Expended
vocabulary vs original vocabulary }
\label{fig:embedding replacement}
\end{figure}

As shown in Figure~\autoref{fig:embedding replacement}, the model with expended vocabulary treats the entire protein ID as a whole when processing protein IDs. In contrast, the tokenizer of the model with original vocabulary will split the protein IDs. The split protein IDs will affect the model's performance in subsequent PPI tasks.

\subsection{The Datasets}
\subsubsection{Human dataset}
The Human dataset contains 4577 unique proteins and 75875 interactions between proteins. The distribution of the Human dataset is shown on \autoref{tab:human}. 
\begin{table}[h!]
\centering
\begin{tabular}{@{}lcc@{}}
\toprule
 Type          & Count & Percentage \\ \midrule
Binding                    & 17977 & 23.70\%    \\
Activation                 & 15470 & 20.41\%    \\
Catalysis                  & 11115 & 14.65\%    \\
Inhibition                 & 10611 & 13.99\%    \\
Expression                 & 9052  & 11.93\%    \\
Post-translational         & 6255  & 8.25\%     \\
Modification and reaction  & 5354  & 7.07\%     \\ \midrule
Total Count   & 75875 &  \\
\bottomrule
\end{tabular}
\vspace{0.3cm}
\caption{Distribution of Human dataset.}
\label{tab:human}
\end{table}

\subsubsection{SHS27K, SHS148K and STRING}
The STRING dataset is a large collection that contains 4,775,154 protein-protein interaction (PPI) records relevant to human biology, covering 15,335 distinct proteins and 572,568 unique interaction events. Two subsets of the STRING database are SHS27k and SHS148k. These subsets are curated by applying specific filters, such as selecting only proteins that are more than 50 amino acids long and exhibit less than 40\% sequence similarity to each other, to ensure diversity and relevance. The SHS27k subset is smaller, with 16,912 PPI entries involving 1,690 proteins and a total of 63,408 interactions. The SHS148k subset is more extensive, containing 99,782 PPI entries, 5,189 proteins, and a high interaction count of 369,041. The distribution of the datasets is shown in \autoref{table:STRING}
\begin{table}[h!]
\centering
\begin{tabular}{@{}lcccccc@{}}
\toprule
& \multicolumn{2}{c}{\textbf{SHS27K}} & \multicolumn{2}{c}{\textbf{SHS148K}} & \multicolumn{2}{c}{\textbf{STRING}} \\
\cmidrule(r){2-3} \cmidrule(lr){4-5} \cmidrule(l){6-7}
\textbf{Type} & \textbf{Count} & \textbf{Percentage} & \textbf{Count} & \textbf{Percentage} & \textbf{Count} & \textbf{Percentage} \\
\midrule

Reaction & 18,162 & 28.65\% & 102,964 & 27.91\% & 1,669,750 & 34.98\% \\
Activation & 7,400 & 11.67\% & 42,516 & 11.52\% & 232,240 & 4.86\% \\
Catalysis & 11,796 & 18.60\% & 67,168 & 18.20\% & 998,266 & 20.91\% \\
Binding & 16,056 & 25.33\% & 93,632 & 25.37\% & 1,610,314 & 33.73\% \\
Ptmod & 2,872 & 4.53\% & 20,153 & 5.46\% & 88,424 & 1.85\% \\
Inhibition & 5,550 & 8.75\% & 34,712 &  9.41\% & 147,676 & 3.09\% \\
Expression & 1,572 & 2.48\% & 7,896 & 2.14\% & 28,484 & 0.60\% \\
\bottomrule
\end{tabular}
\vspace{0.3cm}
\caption{Distribution of SHS27K, SHS148K, STRING dataset.}
\label{table:STRING}
\end{table}

\subsection{Implementation Details and Hyperparameters}
ProLLM is trained on A40-48G. During training, the number of training epochs is 10, the learning rate is 3e-4, the per-device train batch size is 2, the per-device evaluation batch size is 2, the warmup steps are 400, and the weight decay is 0.01.

\subsection{Detail in Embedding Replacement}

To enhance the understanding of protein sequences, we adopt a method that integrates protein sequence vectorization with vocabulary expansion. First,  we query the corresponding protein sequence $S_{\text{p}_{\text{id}}}$ based on the protein's unique identifier $P_{id}$ using the Ensemble BioMart tool. Subsequently, the retrieved protein sequence  $S_{\text{p}_{\text{id}}}$ is fed into the ProtTrans model, which outputs a $1 \times 1024$-dimensional vector $V_p$ encapsulating key information of the sequence. This vector is then used as the embedding vector for the new vocabulary item $P_{id}$ added to the Tokenizer's vocabulary. Through this approach, whenever the model encounters the identifier $P_{id}$, it utilizes the embedding vector $V_p$ generated by ProtTrans for
processing, enabling the model to gain a deeper understanding of protein sequences.

\subsection{Prompt in ProLLM}
\begin{figure}
\centering
\includegraphics[width=1.0\linewidth]{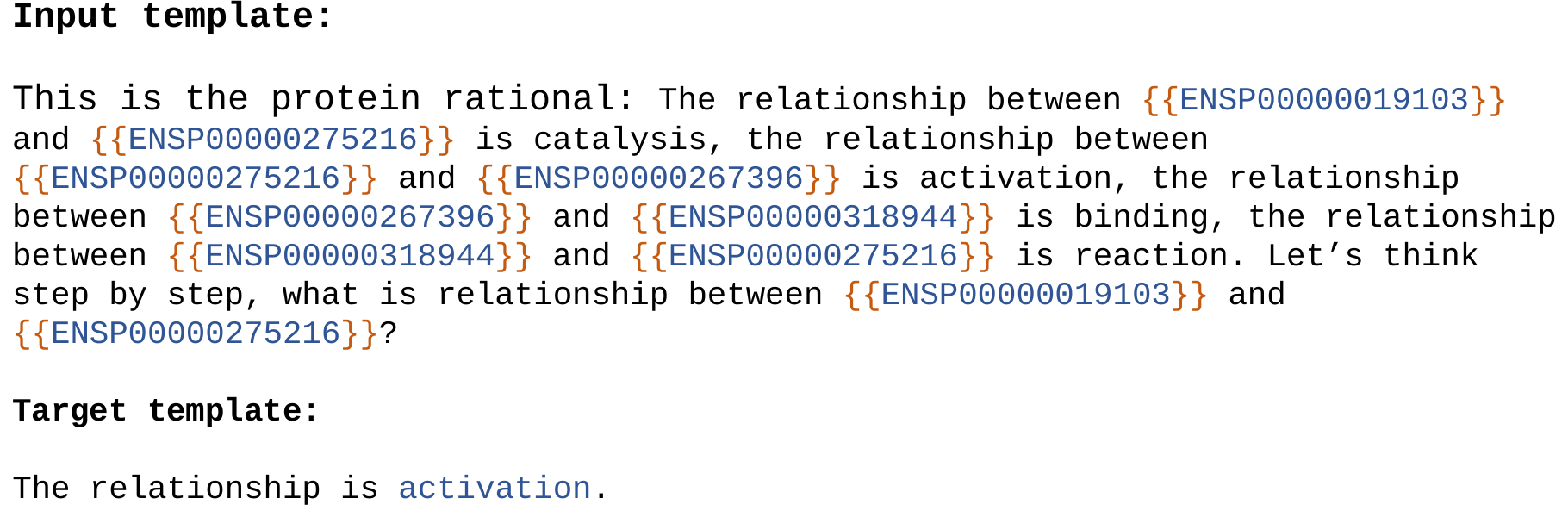}
\caption{One example of ProCoT prompt.}
\label{fig:pcot_prompt}
\end{figure}

\begin{figure}
\centering
\includegraphics[width=1.0\linewidth]{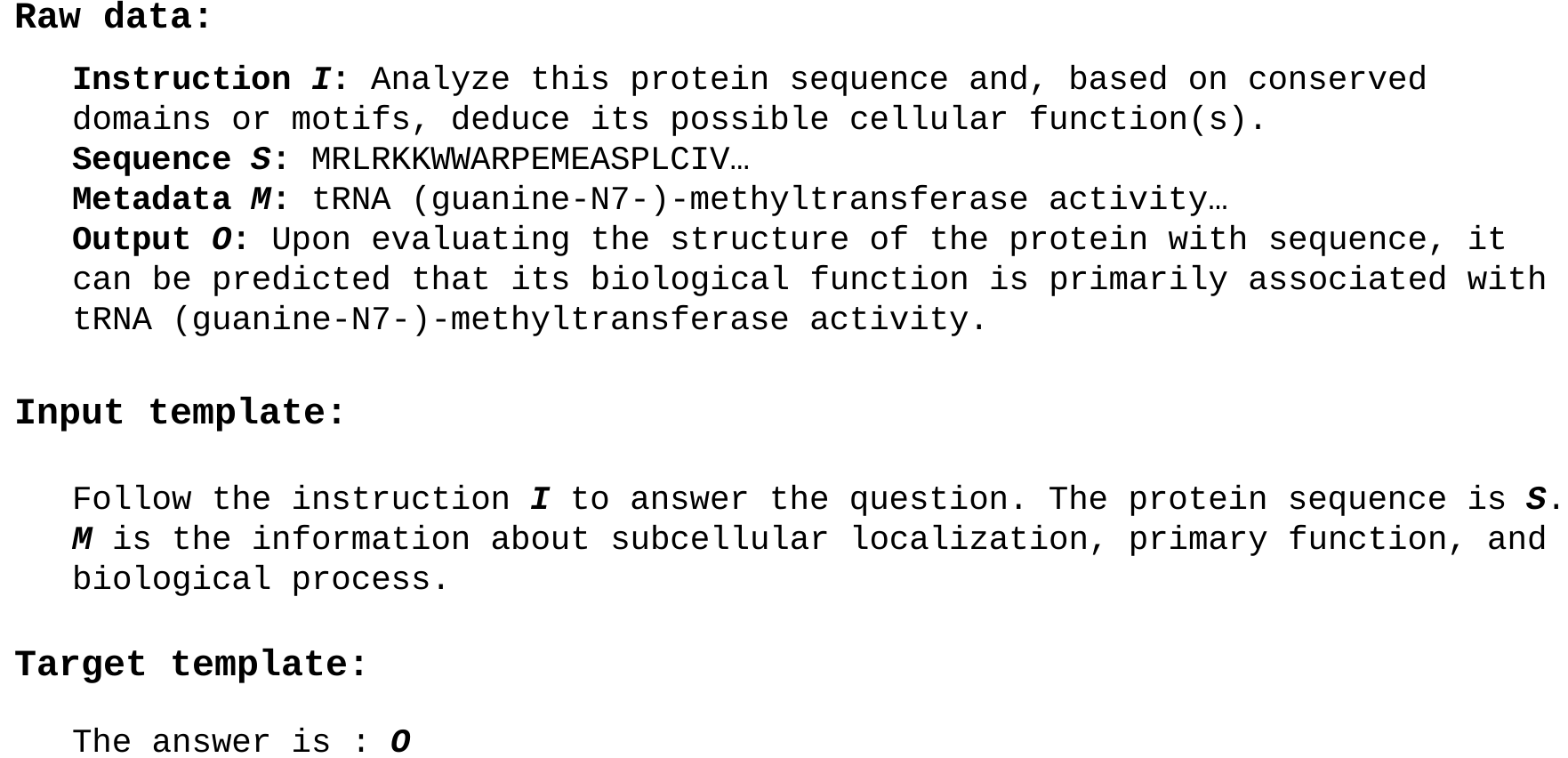}
\caption{Example of instruction fine-tuning prompt.}
\label{fig:instruction}
\end{figure}

The prompt in ProLLM has two type: ProCot prompt and Instruction finetuning prompt. \autoref{fig:pcot_prompt} is an example prompt of ProCot. \autoref{fig:instruction} is the prompt of instruction fine-tuning.

\end{document}